\documentclass[letterpaper,twocolumn,10pt]{article}
\usepackage{usenix-2020-09}

\usepackage{tikz}
\usepackage{amsmath}
\usepackage{url}
\usepackage{amssymb}
\usepackage{multirow}
\usepackage{hyperref}
\usepackage{graphicx}
\usepackage{caption}
\usepackage{subcaption}

\begin{document}

\date{}

\title{\Large \bf Domino: Eliminating Communication in LLM Training via\\Generic Tensor Slicing and Overlapping}


\author{
{\rm Guanhua Wang, Chengming Zhang, Zheyu Shen\thanks{University of Maryland} , Ang Li\footnotemark[1] , Olatunji Ruwase}\\
\\Microsoft DeepSpeed
} 
\maketitle
\begin{abstract}
Given the popularity of generative AI, Large Language Models (LLMs) often consume hundreds or thousands of GPUs for parallelizing and accelerating the training process. Communication overhead becomes more pronounced when training LLMs at scale. To eliminate communication overhead in distributed LLM training, we propose Domino, which provides a generic scheme to hide communication behind computation. By breaking data dependency of a single batch training into smaller independent pieces, Domino pipelines these independent pieces training and provides generic strategy of fine-grained communication and computation overlapping. Extensive results show that, comparing with Megatron-LM, Domino achieves up to 1.3x speedup for LLM training on Nvidia DGX-H100 GPUs.
\end{abstract}

\section{Introduction}

Recent advances in Generative AI (GenAI) enable new application scenarios in various domains, such as chat-bot~\cite{chat-gpt}, text generation and summary~\cite{gpt-3,gpt-4}, image and video content creation~\cite{dalle}. These GenAI applications are based on carefully trained foundation models as large language models (LLMs). Well-establised LLMs are transformer models, such as GPT~\cite{gpt-2, gpt-3,gpt-4} and Llama~\cite{llama,llama-2,llama-3,llama-3-1} series. Given LLM model sizes are usually ranging from tens to hundreds of billion parameters which is far exceeding a single GPU's memory and computation limit, distributed model training over hundreds to thousands of GPUs is necessary. 

For LLM transformer training, three prominent paradigms have emerged: data parallelism (DP), tensor parallelism (TP) and pipeline parallelism (PP). 
Vanilla data parallel training refers to every GPU maintaining a full copy of whole model parameters but training on different input data. Model parameter need to be synchronized at the end of each training iteration among all GPUs in use. 
To mitigate memory pressure from LLMs' huge volume of parameters in DP training, ZeRO~\cite{zero} or FSDP~\cite{fsdp} is often needed to spread model parameters among all GPUs and only recollect the parameters needed to conduct computation. TP and PP belong to model parallelism, where PP~\cite{gpipe,pipe-dream} partitions different model layers on different GPUs. Instead of different GPU holding different model layers, TP~\cite{megatron-lm,megatron-nemo} splits every layer on each GPU, and thus every GPU holds one portion of every model layer. 

Among all distributed model training paradigms, tensor parallelism (TP) has garnered increasing popularity, especially on Nvidia GPUs~\cite{dgx-h100,dgx-a100}. TP was standard practice for single-node multi-GPU training, given its decent system efficiency in high communication bandwidth (i.e., NVlink~\cite{nvlink}, NVSwitch~\cite{nvswitch}) cases. However, because of limited cross-node network bandwidth, TP falls short in multi-node cases. Recently, Nvidia is breaking the bandwidth gap between inter-node and intra-node links. For example, the latest DGX-H100~\cite{dgx-h100} box is equipped with high-bandwidth Infiniband (IB) links with an aggregated bandwidth of 400 GB/s for cross-node communication, which is at the same level of intra-node NVSwitch bandwidth (i.e., 900GB/s on DGX-H100~\cite{dgx-h100}). Therefore, it is time to optimize and propose a TP-only solution for LLM training that covers both single-node and multi-node scenarios. 

\begin{figure}[t]
    \centering
    \includegraphics[width=0.7\linewidth]{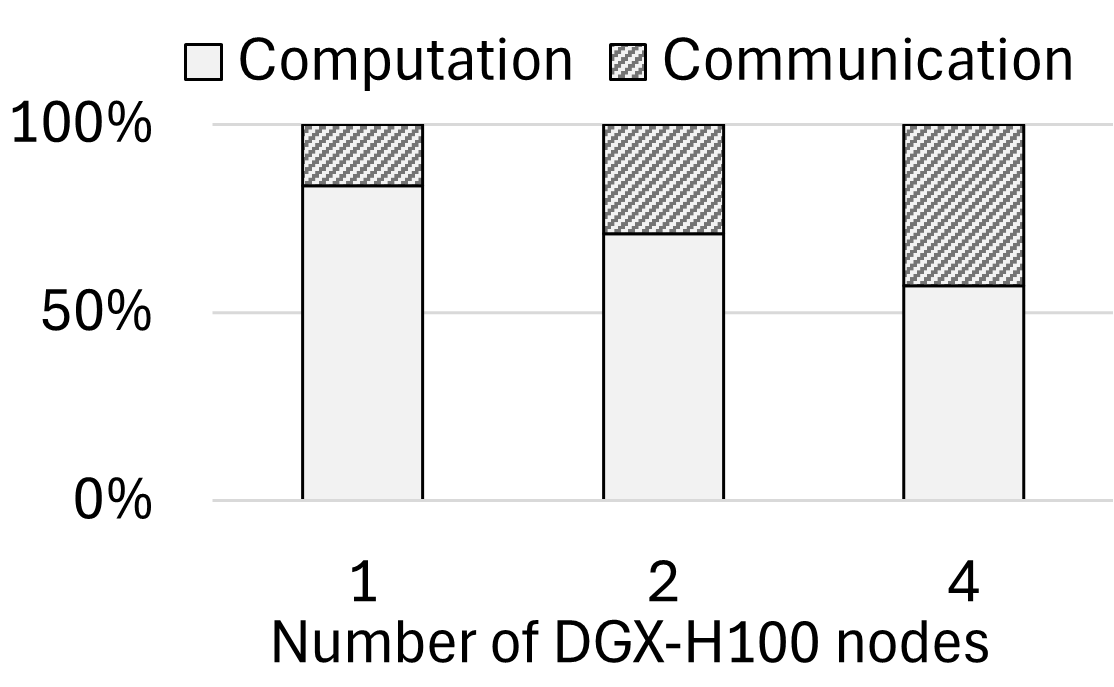}
    \caption{GPT-3-13B computation and communication ratio per training iteration over 1 DGX-H100 node (8 H100), 2 nodes (16 H100) and 4 nodes (32 H100) using TP.}
    \label{fig:gpt-13b-overhead}
\end{figure}

The major overhead of TP is its per-layer global communication, which lies on the critical path of training execution. As described in the literature~\cite{megatron-lm}, every transformer layer needs to communicate twice in forward and another twice in the backward using NCCL collectives~\cite{nccl} (\S ~\ref{sec:tp-comm-overhead}). Given these collective communications happen on the critical path of execution, it is hard to hide these communications behind successive computations with the general communication overlapping strategy used in DP~\cite{zero} or PP~\cite{pipe-dream} training process. Prior arts~\cite{t3-overlap,t3-overhead} report this communication overhead can be up to 45\% of end-to-end training iteration time. As one of our measurements depicted in Figure~\ref{fig:gpt-13b-overhead}, even with the lastest DGX-H100 nodes connected with 400GB/s IB, communication still takes from 17\% to 43\% of end-to-end GPT-3-13B training iteration time. Furthermore, the communication ratio would continue to grow when scale up to more nodes. To mitigate this high communication overhead in TP, prior work~\cite{google-overlap, t3-overlap} provides kernel fusion of a GeMM (General Matrix Multiplication) with its subsequent collective calls (e.g., NCCL~\cite{nccl}) to achieve fine-grained computation-communication overlapping. However, this type of kernel fusion technique limits the overlapping scope and is not general enough to always hide communication behind computation. Especially in the cases where collective communication takes much longer than a single GeMM computation, most of the communication time still stands out as the major training overhead. Furthermore, given that computation on the latest GPUs (e.g., DGX-H100~\cite{dgx-h100}, DGX-B200~\cite{dgx-b200}) is becoming faster, communication overhead is more pronounced in both single node and multi-node cases. 

To provide a generic approach of hiding communication behind computation in TP, we propose Domino, a generic approach that breaks data dependency of transformer model training into pieces, and then pipelines these pieces training to overlap communication with computation. Besides traditionally TP can only be used within a node, Domino provides a uniformed TP solution for both single-node multi-GPU and multi-node multi-GPU cases. Compared with previous GeMM+NCCL fusion solutions, Domino provides a much wider scope of computation and communication overlapping (e.g., AllReduce not only overlaps with a single GeMM, but also LayerNorm, DropOut, etc). Additionally, any kernel fusion and optimization techniques can be easily integrated with Domino as a drop-in replacement to further boost overall system efficiency. 

Extensive benchmark results are collected from Nvidia latest hardware DGX-H100 boxes, which are connected with 3200 Gb/s (i.e., 400 GB/s) InfiniBand (IB) fabrics~\cite{dgxh100-ib}. We benchmark training with popular transformer models such as GPT-3~\cite{gpt-3} and Llama-2~\cite{llama-2}. Compared with state-of-the-art TP implementation Megatron-LM from Nvidia, Domino achieves up to 1.3x speedup for both single-node and multi-node cases. Overall, Domino provides a generic approach of flexible overlapping of communication with a wide range of computation kernels for transformer training. 

We summarize our key contributions as follows:

\begin{itemize}
\item To the best of our knowledge, Domino is the first work providing an end-to-end solution of generic communication-computation overlapping for tensor-parallelism-only training in both single-node and multi-node cases.
\item Compared with previous arts, Domino provides a more flexible and wider range of computation and communication overlapping strategies.
\item Experiment results on DGX-H100 boxes show that, compared with Megatron-LM, Domino achieves up to 1.3x speedup for GPT and Llama models training. 
\item Domino will be open-sourced and released as part of \url{https://github.com/microsoft/DeepSpeed}
\end{itemize}

\section{Background and Motivation}
In this section, we first describe the most widely used distributed transformer training schemes as data parallelism (DP), tensor parallelism (TP) and pipeline parallelism (PP) as \S~\ref{sec:dist-train-scheme}. Then we illustrate why TP is becoming increasingly popular among these three approaches as \S~\ref{sec:tp-trending}. Finally, we analyze the communication overhead of TP in both single-node multi-GPU and multi-node multi-GPU cases in \S~\ref{sec:tp-comm-overhead}.

\subsection{Distributed Training Schemes}
\label{sec:dist-train-scheme}
There are mainly three different kinds of paradigms for distributed LLM training, namely, data parallelism (DP), tensor parallelism (TP), and pipeline parallelism (PP). 

In vanilla DP, each GPU maintains a full copy of model weights and consumes different input data. At the end of each training iteration, all GPUs involved conduct an AllReduce operation to synchronize model parameters. Specifically for transformer models, ZeRO~\cite{zero} and FSDP~\cite{fsdp} are widely used to reduce memory pressure on devices. In these fully shared data parallel schemes, whole model weights are often evenly split across all GPUs. When computation needs to happen on a specific layer, every GPU recollects the full weights of this particular layer by conducting an AllGather operation among all GPUs for this layer. Once the computation is done, every GPU releases the whole layer weights and only maintains the portion of weights that were originally assigned on each GPU. Therefore, ZeRO/FSDP can be regarded as a memory efficient data parallelism scheme that trades more communication with less on-device memory footprint. 

PP and TP are both representative of model parallelism techniques. PP partitions a layer or a group of layers on a single GPU and then pipeline executes from GPU holding the first layer to GPU holding the last layer during forward propagation, and then backward propagation in the reverse order. 
Compared with PP, TP partitions the model in an orthogonal direction, where each GPU holds a portion of every model layer, such that every GPU can compute from the first layer to the last layer by itself without blocking caused by sequential dependency created among PP stages~\cite{gshard}. 
Additionally, TP seems to have a similar model partition strategy as ZeRO/FSDP. The main difference here is, compared with ZeRO/FSDP, TP never recollects weights during forward or backward computation but synchronizes on activations or gradients via AllReduce. Compared with DP and PP, TP provides the highest system efficiency or training throughput with high bandwidth communication links~\cite{pytorch-tp, pytorch-tp-roadmap, vllm-tp}. 

\begin{figure}[t]
    \centering
    \includegraphics[width=1\linewidth]{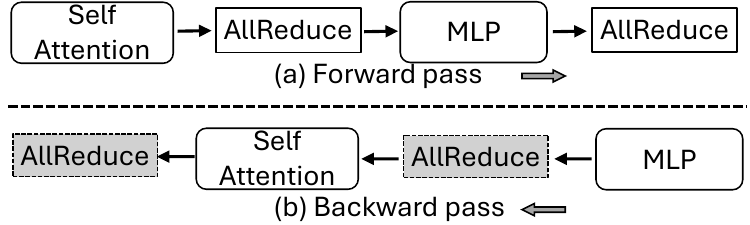}
    \caption{4 AllReduce in each transformer block in TP training. Two blank AllReduce boxes are in forward pass, and the other two grey AllReduce boxes are in backward pass.}
    \label{fig:tp-ar}
\end{figure}

\subsection{TP is Trending}
\label{sec:tp-trending}
Tensor parallelism is gaining popularity for LLM workloads on Nvidia GPUs. AI practitioners have recently witnessed substantial improvements in both TP software and hardware stacks.

On the software side, Nvidia consistently enhances its Megatron-LM~\cite{megatron-lm,megatron-nemo} software stack as the state-of-the-art TP implementation. Megatron-LM achieves greater efficiency through integrating with more fine-tuned and customized compute kernels sourced from libraries like apex~\cite{apex}, cutlass~\cite{cutlass}, cublas~\cite{cublas}, cudnn~\cite{cudnn}. Besides that, Megatron-LM also involves new features to enhance overall system throughput, including selective activation checkpointing, and sequence parallelism strategies~\cite{megatron-nemo}.

More importantly on the hardware front, Nvidia is pushing hard to bridge the bandwidth gap between intra-node and cross-node links, which is essential for extending TP to cross-node use cases. The latest DGX-H100 node is equipped with eight Nvidia ConnectX-7 InfiniBand (IB) cards~\cite{nv-connectx7}, and each provides 400 Gb/s bandwidth. Thus each DGX-H100 box achieves 400 GB/s cross-node communication bandwidth, which is comparable to intra-node NVLink/NVSwitch bandwidth as 900 GB/s~\cite{dgx-h100}. 
Furthermore, advancements in Nvidia's network infrastructure suggest that future DGX systems could potentially integrated with Nvidia ConnectX-8 IB cards~\cite{nv-connectx8}, offering up to 800 GB/s aggregated cross-node bandwidth, approaching the bandwidth available with intra-node NVLink/NVSwitch connections. 

With these advancements in software and hardware, both PyTorch~\cite{pytorch} and raising vLLM~\cite{vllm} communities lean towards applying TP for both transformer training and inference. For instance, the PyTorch team sets TP as one major future direction of efficient LLM training in their recent release~\cite{pytorch-tp}, and also sets improving TP scalibility as Key Results (KR) in Meta PyTorch team 2024 H2 roadmaps~\cite{pytorch-tp-roadmap}. Similarly, on the inference side, vLLM has embraced TP as the only option for distributed LLMs serving~\cite{vllm-tp}. 

Given the growing popularity of TP and recent breakthroughs in IB hardware, it is now imperative to establish a uniformed LLM training solution of TP for both single-node and cross-node scenarios. Before delving into our Domino design, we next discuss the communication overhead associated with TP.

\begin{figure}[t]
    \centering
    \includegraphics[width=1\linewidth]{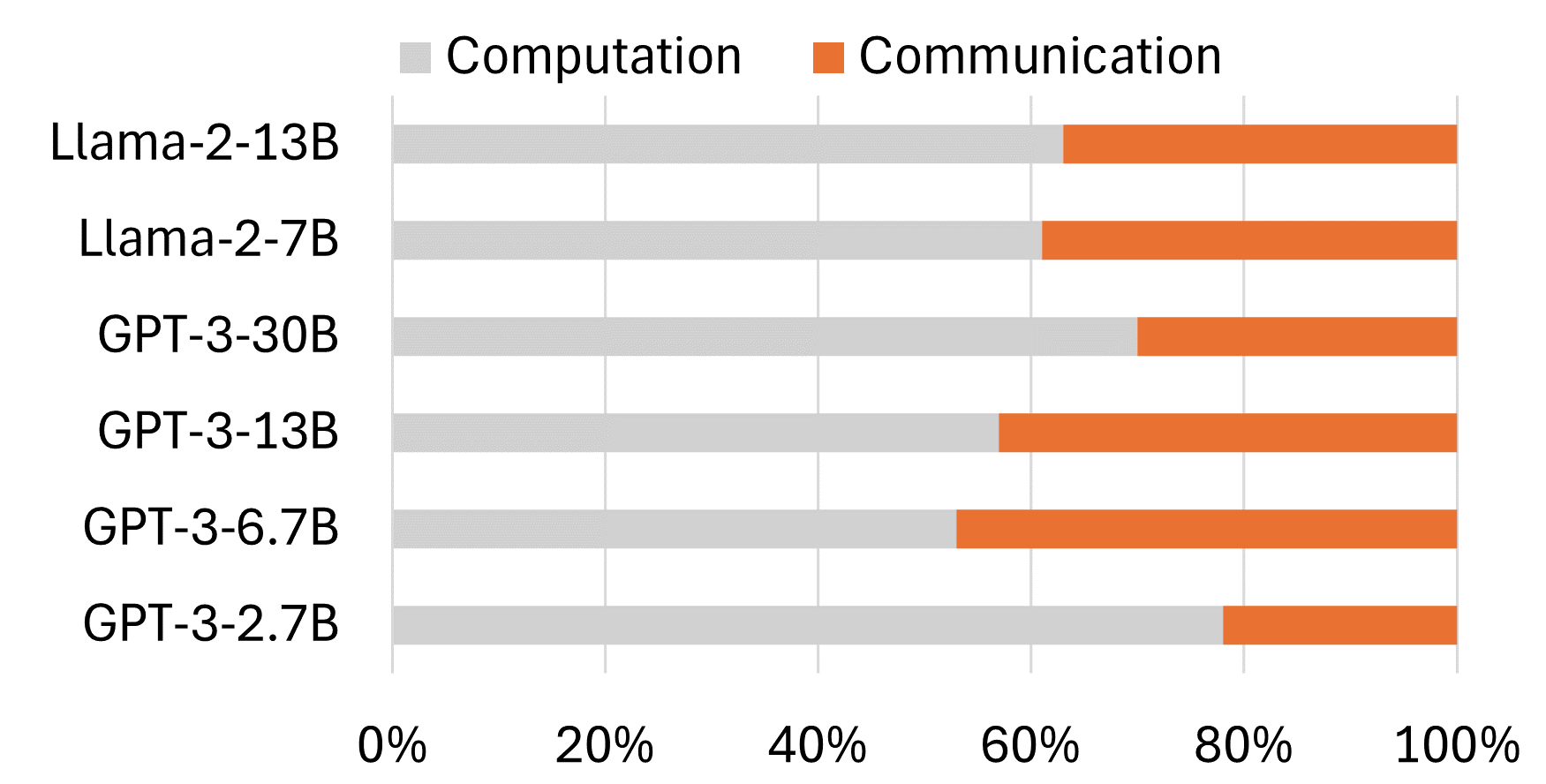}
    \caption{TP computation and communication ratio per training iteration on varied model types and model sizes over 1 to 4 DGX-H100 nodes (8 to 32 H100 GPUs).}
    \label{fig:tp-comm-overhead}
\end{figure}

\subsection{TP Communication Overhead}
\label{sec:tp-comm-overhead}

We conduct measurements using state-of-the-art TP implementation from Nvidia as Megatron-LM~\cite{megatron-lm,megatron-nemo, megatron-code}.

TP communication possesses more unique characteristics compared with PP or DP solutions, mainly because \emph{it resides on the critical path of every input batch training}. Hiding communication behind computation is standard practice not only limited to LLM training but also extensively applied in all distributed system environments~\cite{mpi-overlap,codes-overlap,dcuda-overlap, wavelet,multi-resource-overlap}. 
For transformer training using DP or PP, the overlapping of communication and computation is quite straightforward since we can schedule communication on a side channel thus bypassing the critical execution path. 
For DP approaches like ZeRO~\cite{zero} or FSDP~\cite{fsdp}, pre-fetching weights enable overlap with computation, as weights inherently do not have any sequential data dependency. 
PP naturally overlaps communication and computation by processing on different input batches. For instance, on each GPU, PP can overlap the previous batch's communication with computation for current batch data.  

As described in Megatron-LM~\cite{megatron-code,megatron-lm}, each transformer block comprises a self-attention layer and an MLP (multi-layer perceptron) layer.
As shown in Figure~\ref{fig:tp-ar}, both self-attention and MLP layers trigger an AllReudce operation in both forward and backward propagation. Consequently, each transformer block necessitates a total of 4 AllReduce per each training iteration. Given a language model consisting of $N$ stacked transformer blocks, this results in $4\times N$ AllReduce per iteration, imposing significant communication overhead. Furthermore, as discussed above, traditional methods fail to hide this communication behind computation, thereby placing all TP communication overheads on the execution critical path.

We measure the communication overhead in TP training with Megatron-LM across GPT-3 and Llama-2 model series with different model sizes. The models run on different numbers of DGX-H100 nodes ranging from 1 to 4 (i.e., 8 to 32 H100 GPUs) depending on model sizes and batch sizes. As shown in Figure~\ref{fig:tp-comm-overhead}, the communication time is ranging from 22\% to 47\% of end-to-end iteration time. This finding underscores that, despite the utilization of high bandwidth NVLink/NVSwitch/Infiniband interconnects, the communication overhead remains a significant portion of training iteration time. This is primarily due to more significant increase in computation power per each GPU (e.g., H100) compared with previous generations (e.g., V100, A100), thereby making communication overhead still stand out. 



\section{Domino Design}
\label{sec:design}
In this section, we describe the detailed design of Domino architecture. We first provide an overview of system architecture (\S~\ref{sec:overview}). Then we detail how to generically partition computation and overlap sequences of computation kernels with communication (\S~\ref{sec:row-split},\S~\ref{sec:col-split}, \S~\ref{sec:hybrid-split}).

\begin{figure}[t]
    \centering
    \includegraphics[width=1\linewidth]{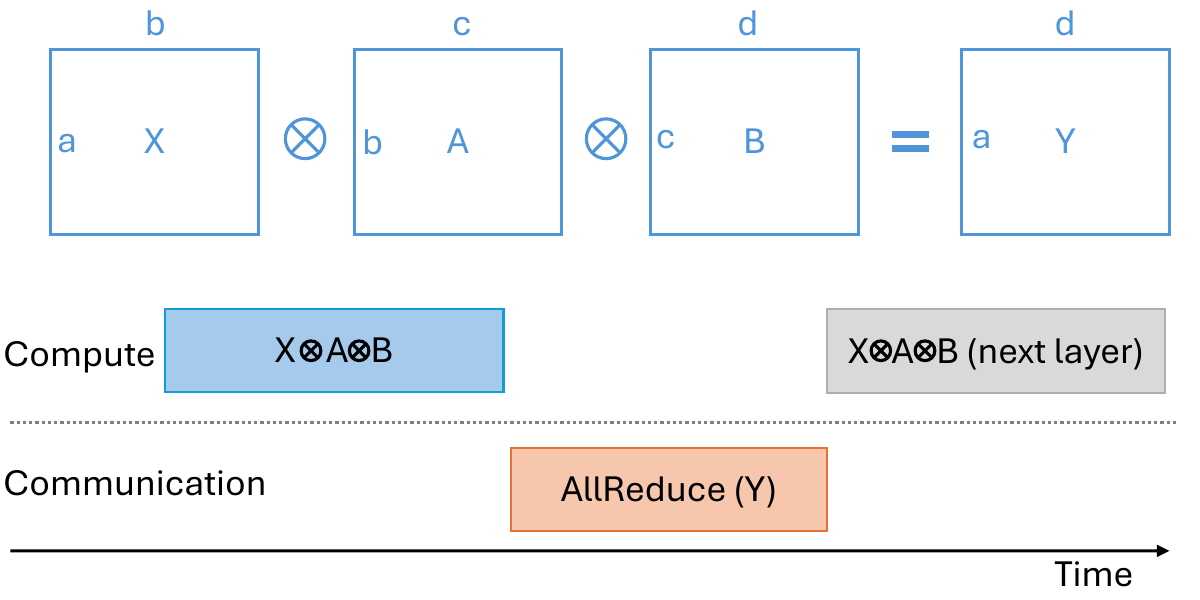}
    \caption{Forward pass of single Self-Attention / MLP layer.}
    \label{fig:design-base}
\end{figure}

\subsection{Overview}
\label{sec:overview}
We first describe overall workflow of Domino. Given standard transformer architecture, we abstract both self-attention layer and MLP (multi-layer perception) layer as weight tensors of $A\_FULL$ and $B\_FULL$, where $A\_FULL$ stands for attention weights (i.e., $W_q$,$W_k$,$W_v$) for self-attention layer but linear weights for MLP, and $B\_FULL$ is linear weights for both self-attention and MLP layer. For ease of illustration, we describe our partition strategy in forward propagation since backward propagation is just in reverse execution order. Given layer input data $X$, both self-attention and MLP layers' computation can be abstracted as Equation~\ref{eq:no-split}.

\begin{equation}
\label{eq:no-split}
 X \otimes A\_FULL \otimes B\_FULL = Y\_FULL   
\end{equation} 


As shown in Figure~\ref{fig:design-base}, TP (e.g., Megatron-LM) splits whole weights tensor $A\_FULL$ \emph{column-wise} as set$\{A\ |\ A_i\ on \ GPU_i\}$, and weights tensor $B\_FULL$ \emph{row-wise} as set $\{B\ |\ B_i\ on \ GPU_i\}$ for both self-attention and MLP layers. After every GPU getting its own $A$ and $B$ weights partitions, TP executes $X \otimes A \otimes B = Y$ and then conducts AllReduce on set $\{Y\ |\ Y_i\ on \ GPU_i\}$ sequentially to recover $Y\_FULL$, which makes communication overhead completely stand-out. 

To hide TP communication behind computation, Domino provides extra and generic tensor partition in two dimensions on every GPU: \emph{row-wise} split on inputs $X$ and \emph{column-wise} split on weights $B$ on top of original TP model partitions. 







At high level, Domino generically breaks TP's $X \otimes A \otimes B$ into smaller compute units without data dependency. Then we pipeline these independent compute units with collective communication to achieve fine-grained computation and communication overlapping. With the latest trend that attention computation is modularized and highly optimized like flash-attention~\cite{flashattention, flashattention2}, windowed-attention~\cite{window-attn}, etc., we keep $A$ untouched and do not conduct any tensor partitioning on $A$. Therefore, we only conduct tensor slicing on input tensor $X$ (\S~\ref{sec:row-split}) and the second group of linear weights as $B$ (\S~\ref{sec:col-split}). 
We also provide a hybrid tensor partition strategy of both $X$ and $B$ (\S~\ref{sec:hybrid-split}). After these tensor slicing, Domino breaks $X \otimes A \otimes B$ into pieces and removes data dependency. Then we enable computation-communication overlapping on these independent pieces to reduce communication overhead in TP.

Prior to real model training, we benchmark system efficiency to determine the tensor partition granularity with grid search. These benchmarks guide our selection of tensor computation sizes to ensure minimal impact on computation kernel efficiency. To further enhance system efficiency, Domino also adopts the latest features like kernel fusion, torch.compile~\cite{torch-compile} and CUDAGraph~\cite{cuda-graph,nv-cuda-graph} techniques from PyTorch and Nvidia as described in \S~\ref{sec:kernel-opt}.

\subsection{Row Split on Inputs}
\label{sec:row-split}
\begin{figure}[t]
    \centering
    \includegraphics[width=1\linewidth]{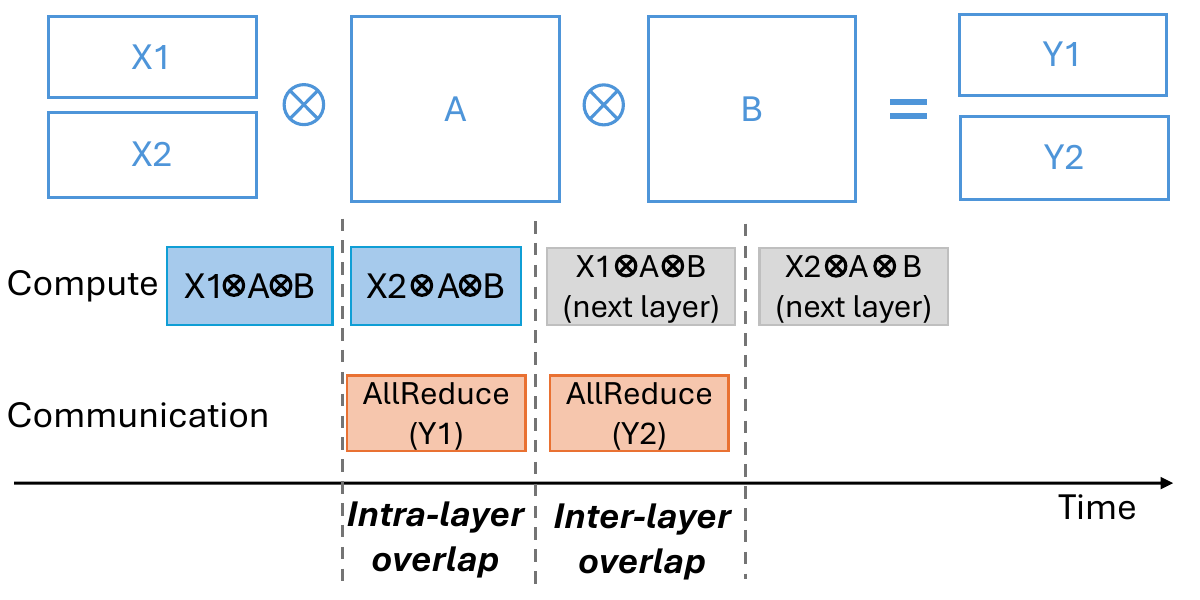}
    \caption{Domino row-wise (batch-dim) split on input $X$.}
    \label{fig:design-row}
\end{figure}

We first discuss row-wise split on input data, which refers to tensor partitioning on $X$ in \S~\ref{sec:overview}. For ease of illustration in Figure~\ref{fig:design-row}, we simplify and assume all tensors are with 2 dimensions.
Then $X$ can be split in either row dimension or column dimension. In reality, each layer's input tensor usually is 3D as $(batch, seq, hidden)$. We map row/column dimensions of our example 2D tensor into batch/hidden dimensions of real 3D tensor, respectively.

If we split input in column dimension to $N$ chunks, the communication volume will be $N^2$ times bigger than vanilla baseline. As shown in Figure~\ref{fig:design-base}, assuming $X$ with tensor shape as $(a,b)$, $A$ as $(b,c)$ and $B$ as $(c,d)$. If we do a column-wise split on X and shard it as $[X_1,X_2...X_N]$ with each shape of $(a, b/N)$, we will get $N^2$ output tensors with original $Y$ shape $(a,d)$ after $X \otimes A \otimes B = Y$ with proper reshaping on $A$.  
To avoid this communication volume blow-up, we choose row-wise split $X$ to $(a/N,b)$ in Figure~\ref{fig:design-row}, which refers to input tensor partition ($X1,X2$) on batch dimension in reality. 

\begin{equation}
\label{eq:x-a}
X \otimes A = \left\{ \begin{array}{rcl}
 softmax(\frac{(X*W_q)(X*W_k)^T}{\sqrt{d_k}})(X*W_v) & \mbox{for} & Attn \\
X*A & \mbox{for} & MLP
\end{array}\right.    
\end{equation} 

Note that our row-wise split happens on input's batch dimension, it is mathematically equivalent to vanilla baseline. Given row (batch-dim) split on $X$ mainly affects abstracted computation of $X\otimes A$, we illustrate in details on $X\otimes A$ to show equivalence of our row-split on $X$ and baseline. 
Element-wise operations like GeLU() and dropout() are completely independent along batch dimension of $X$, we exclude them for simplicity.
Then we get simplified $X\otimes A$ as Equation~\ref{eq:x-a}. 

For MLP, $X \otimes$ A is just GeMM between $X$ and $A$. Therefore, as a toy example in Figure~\ref{fig:design-row}, row-wise split on $X$ is equivalent to baseline as Equation~\ref{eq:xa-mlp}.
\begin{equation}
\label{eq:xa-mlp}
\left[ \begin{array}{c} X1 \\ X2 \end{array} \right] * A = X * A
\end{equation} 
For self-attention, we abstract it as $softmax(f(X))g(X)$. For the second part $g(X)=X*W_v$, the equivalence proof here is the same as Equation~\ref{eq:xa-mlp} since it is just a GeMM operation. For $f(X)=\frac{(X*W_q)(X*W_k)^T}{\sqrt{d_k}}$, its output dimensions are $(batch, seq, seq)$. Since Softmax() is conducted on the last dimension of $f(X)$ output as sequence-dim, which is completely independent of first batch dimension. 
Since $softmax(f(X))$ and $g(X)$ are both independent in batch dimension and their product is also independent in batch dimension, 
row-wise split on $X$ for self-attention layer is also equivalent to baseline without tensor slicing. 



\textbf{Data dependency:} Since the batch dimension of input tensor is completely independent, no synchronization is needed across all transformer layers. As depicted in Figure~\ref{fig:design-row}, Domino's row-split on input achieves both \emph{intra-layer} and \emph{inter-layer} (i.e., overlap communication with successive layer computation) computation and communication overlapping. 

\begin{figure}[t]
    \centering
    \includegraphics[width=1\linewidth]{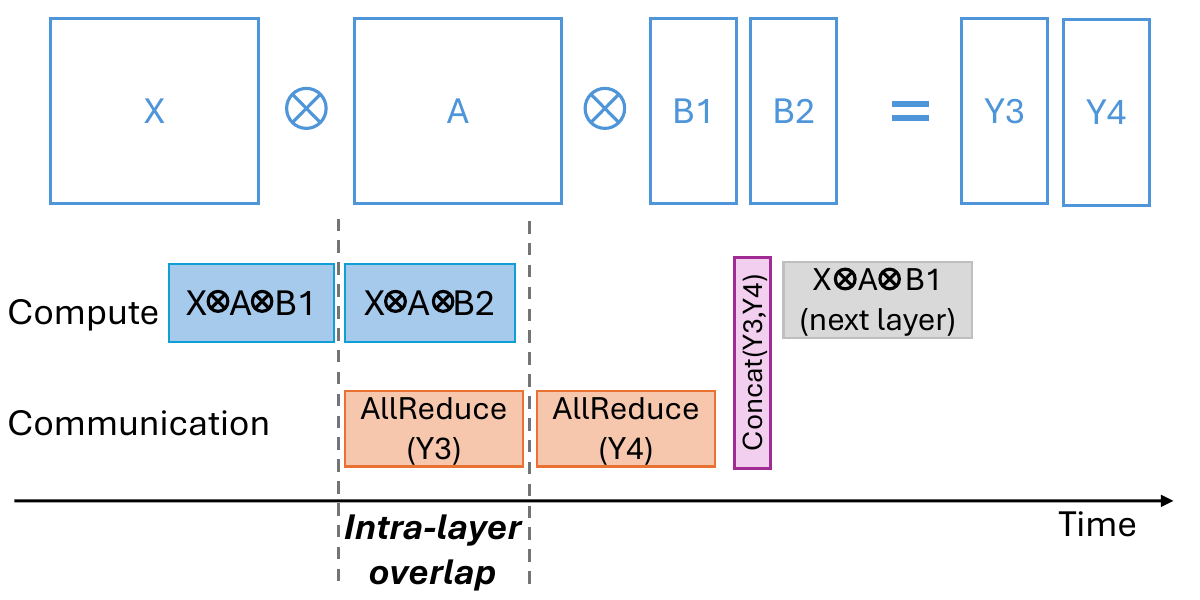}
    \caption{Domino col-wise (last-dim) split on weights $B$.}
    \label{fig:design-column}
\end{figure}

\subsection{Column Split on Weights}
\label{sec:col-split}
With similar analysis as \S~\ref{sec:row-split}, partitioning weights tensor $B$ in row-dimension for $N$ partitions will lead to $N^2$ times communication volume blow-up. To avoid this, we split the weight tensor $B$ on the column dimension to keep the communication volume the same as the vanilla baseline.

As shown in Figure ~\ref{fig:design-column}, for $B$, we split it column-wise to $N$ partitions, and each partial output will have the shape of $(a, d/N)$. After collecting all $N$ chunks, we concatenate these partial results ((e.g., Concat(Y3,Y4) in Figure~\ref{fig:design-column})) at the end of each $X \otimes A \otimes B$ layer computation. 

Now we prove column-wise split on weights $B$ is equivalent to baseline without tensor partition. Since dropout() happens after our concatenation as concatenation output is identical to baseline, it can be safely removed from our proof domain. By excluding element-wise dropout() operation, $(XA) \otimes B$ is just GeMM for both self-attention and MLP layers. Thus, the equivalence is shown as Equation~\ref{eq:xa-b}.

\begin{equation}
\label{eq:xa-b}
(XA)\otimes B = (XA) \otimes [B1, B2]
\end{equation} 


\textbf{Data Dependency}: given this column-wise split on B, for both self-attention layer and MLP layer, the computation output needs to be synchronized at the end of layer execution. As a toy example of column-wise split of 2 shown in Figure~\ref{fig:design-column}, Domino achieves \emph{intra-layer} computation communication overlapping but needs synchorize (i.e., Concat(Y3,Y4) in Figure~\ref{fig:design-column}) before moving to next self-attention or MLP layer computation.


\subsection{Hybrid Split}
\label{sec:hybrid-split}
For extremely large LLMs~\cite{gpt-4}, we provide a hybrid model split on both input $X$ and last weight tensor $B$. This hybrid solution is necessary since either row-split or column-split alone would cause narrow shape tensor which is impossible to achieve good computation efficiency. After doing row-wise split on $X$ together with column-wise split on $B$, Domino can achieve super fine-grained computation and communication overlapping. The aggregated communication size of $X \otimes A \otimes B$ still remains the same as vanilla baseline.  

\textbf{Data Dependency}: Inherited from column-wise split on $B$, for both self-attention layer and MLP layer, final computation outputs need to be synchronized column-wise (i.e., Concat(Y3,Y4) in Figure~\ref{fig:design-column}) but non-blocking row-wise. Therefore, the hybrid split can only achieve \emph{intra-layer} computation and communication overlapping. 

\section{Implementation}
\label{sec:implementation}
We now discuss implementation details, which includes row-wise partitioning strategy for input data (\S~\ref{sec:implement-input}), column-wise partitioning approach for model weights (\S~\ref{sec:implement-weight}), and further optimization on computational kernels (\S~\ref{sec:kernel-opt}). 

\subsection{Tensor Partitioning on Inputs}
\label{sec:implement-input}

\begin{figure}[t]
    \centering
    \includegraphics[width=1.0\linewidth]{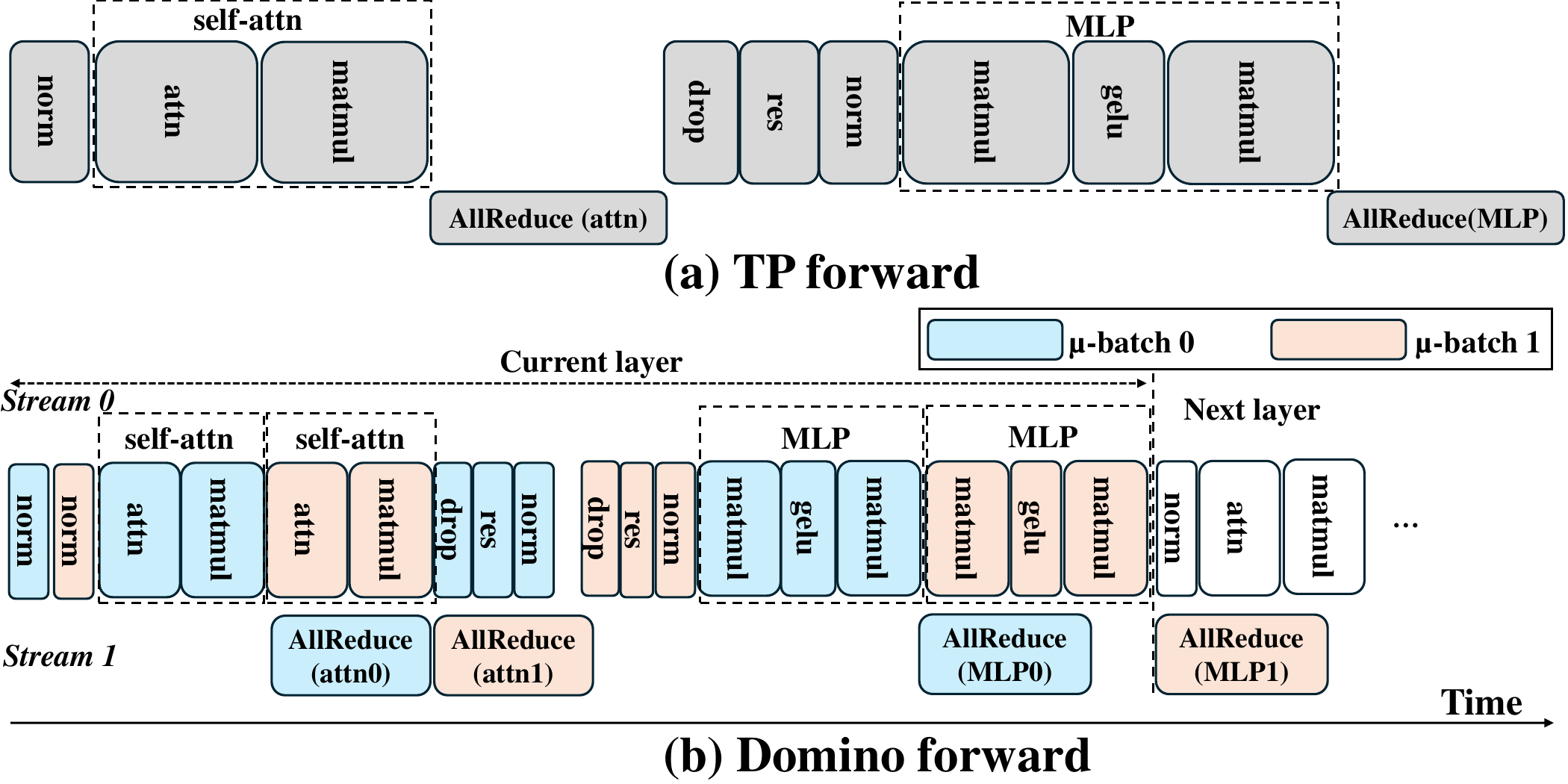}
    \caption{Transformer block (i.e., 1 self-attn and 1 MLP) forward phase. Upper figure is vanilla TP implementation, and bottom figure is Domino implementation.}
    \label{fig:implement-fwd}
\end{figure}

We first illustrate the implementation of our novel input partitioning in both forward and backward propagations, separately.

\subsubsection{Forward phase}
\label{sec:implement-input-fwd}
Users can define the number of partitions $p_1$ for the input, after which the input is divided into $p_1$ partitions along the batch dimension. 
A for-loop iterates through each partitioned $\mu$-batch sequentially. Figure \ref{fig:implement-fwd} depicts the forward phase of a simple example where the layer input is split into two $\mu$-batches (i.e., $p_1 = 2$). 

\begin{figure}[t]
    \centering
    \includegraphics[width=1.0\linewidth]{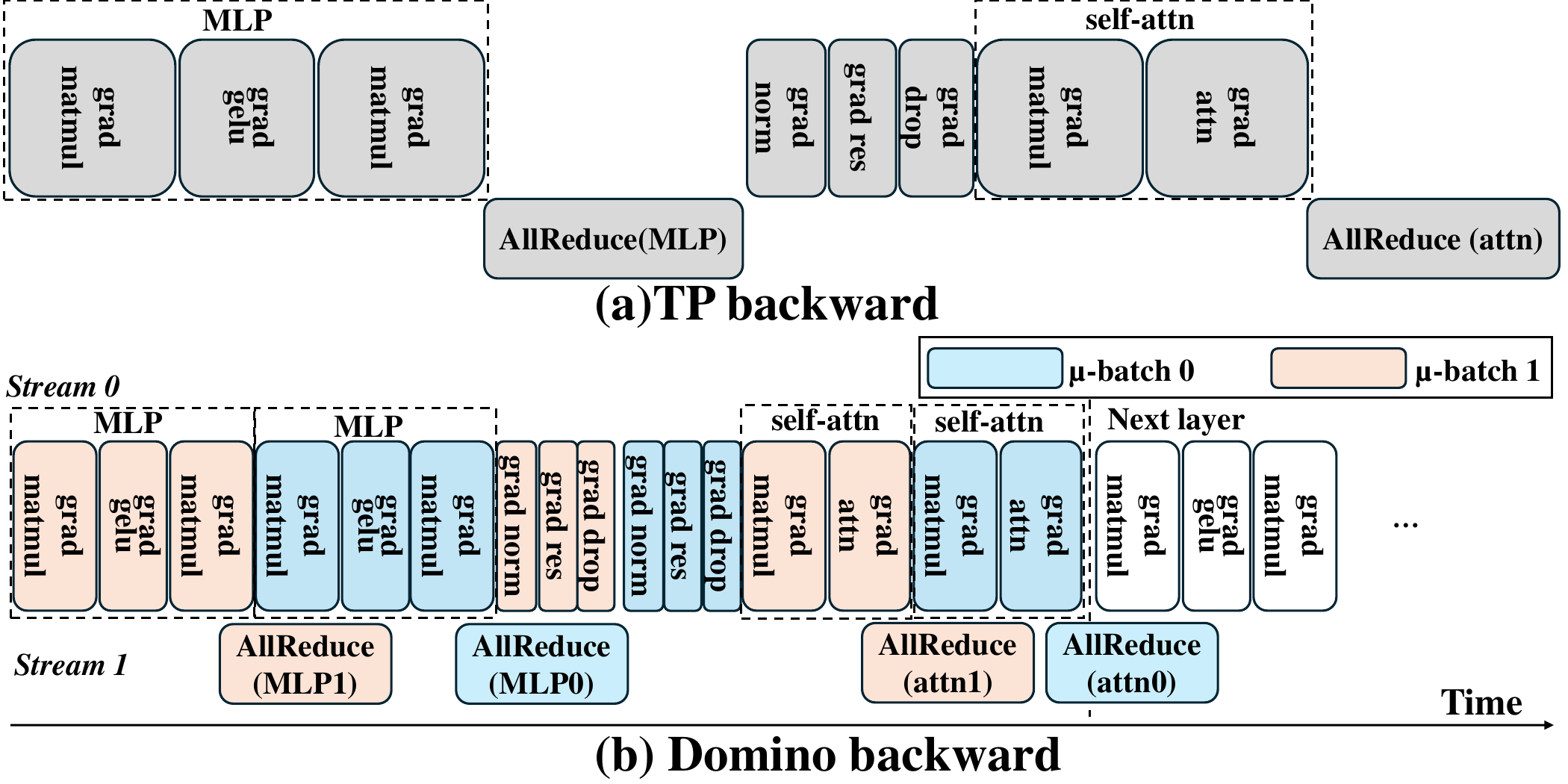}
    \caption{Transformer block (1 self-attn and 1 MLP) backward phase. Upper figure is vanilla TP implementation, and bottom figure is Domino implementation.}
    \label{fig:implement-bwd}
\end{figure}

In Figure~\ref{fig:implement-fwd}(a), to hide AllReduce communication (i.e., \emph{AllReduce (attn)} in Fig.~\ref{fig:implement-fwd}(a)) after the self-attention layer, we first execute the self-attention of $\mu$-batch 0 and then initiate its AllReduce (i.e., \emph{AllReduce(attn0)} in Figure~\ref{fig:implement-fwd}(b)) asynchronously to prevent GPU blocking on communication. Subsequently, we immediately proceed self-attention on $\mu$-batch 1. The communication of self-attention of $\mu$-batch 1 (i.e., \emph{AllReduce(attn1)}) overlaps with layer normalization, residual, and dropout operations.
The reason for grouping multiple $\mu$-batches' dropout, residual, layerNorm not only enables hiding \emph{AllReduce(attn1) in Figure~\ref{fig:implement-fwd}(b)} for forward pass, but also provides proper overlapping space for \emph{Allreduce(MLP0)} in backward pass shown in Figure~\ref{fig:implement-bwd}(b).

Similarly to hide \emph{AllReduce(MLP0)} communication in Figure~\ref{fig:implement-fwd}(b) in MLP forward, we initiate this AllReduce after MLP computation on $\mu$-batch 0 asynchronously, enabling immediately execute MLP of $\mu$-batch 1 to achieve overlapping. Additionally, \emph{AllReduce(MLP1)} after MLP of $\mu$-batch 1 will overlap with the computation of $\mu$-batch 0 in the successive transformer block.

\subsubsection{Backward phase}
\label{sec:implement-input-bwd}
The corresponding backward is mostly generated by torch.autograd(). Figure~\ref{fig:implement-bwd} shows a toy example of backward pass where the input hidden states are split into two $\mu$-batches ($p_1 = 2$). We carefully organize the execution of gradient computation in these $\mu$-batches to overlap gradient computation and communication. 

In Figure~\ref{fig:implement-bwd}, we first adopt similar cross $\mu$-batch computation and communication overlapping as described in \S~\ref{sec:implement-input-fwd}. 
To further broaden overlapping scope, we also adopts overlapping communication with weights gradient computation within the same $\mu$-batch. For example, \emph{AllReduce(MLP1)} in Figure~\ref{fig:implement-bwd}(b) partially overlaps with \emph{grad matmul} computation of its own $\mu$-batch 1 (i.e. 3rd orange block from left). Each \emph{grad matmul} usually involves two separate kernel computation as inputs gradient and weights gradient computation. This sub-module overlapping can be achieved by first calculating inputs gradient inside 2nd \emph{grad matmul} in MLP layer of $\mu$-batch 1 (i.e. 3rd orange block from left), and then trigger its weights gradient computation and inputs gradient communication simultaneously. 

However, accurate control of gradient communication behavior to overlap with gradient computation is challenging because PyTorch automatically generates the gradient computation graph~\cite{pytorch}. To precisely control communication start/end time, our initial attempt to manually implement customized backward pass leads to poor throughput performance due to triggering less efficient kernels than torch.autograd(). 
To tackle this issue, we developed a \emph{no-operation} module. This module receives the communication handle during the forward phase and retains it for use during the backward phase. Our \emph{no-operation} module integrates seamlessly with torch.autograd(). This approach allows us to precisely control the completion time of asynchronous communication without complicated code modification.


To sum up, \emph{Domino enables up to $\sim$100\% communication hiding behind computation with our batch-split on inputs.} 


\subsection{Tensor Partitioning on Weights}
\label{sec:implement-weight}
Users can also define the number of partitions $p_2$ for weights. Subsequently, $p_2$ linear modules are initialized, each with hidden dimension scaled by $\frac{1}{p_2}$.

Bottom half of Figure~\ref{fig:design-column} shows a toy example of the weight partition where the number of partitions for weights is 2. Specifically, we first execute the first linear module (i.e., $X\otimes A \otimes B1$) to generate the first half result (i.e., $Y3$). We then trigger asynchronous non-blocking AllReduce on the first half result.
After that, we immediately execute the second half linear module ($X\otimes A \otimes B2$). Therefore, \emph{AllReduce(Y3)} is overlapped with $X\otimes A \otimes B2$. In the backward, we adopt similar sub-module overlapping strategy as discussed in \S~\ref{sec:implement-input-bwd}. 

An obstacle here is to fully restore hidden dimension (i.e. \emph{concat(Y3,Y4)} in Figure~\ref{fig:design-column}) for subsequent operations (e.g., layerNorm, dropout, etc.). torch.cat() often allocates GPU memory more than needed~\cite{page-attn}, which may trigger unnecessary OOM  (out-of-memory) errors. To achieve concatenation on hidden dimension without torch.cat(), we pre-allocate a big buffer to store the first half (i.e., \emph{Y3}) and the second half (i.e., \emph{Y4}) result sequentially in Figure~\ref{fig:design-column}. However, this method still incurs extra memory copy (MemCpy) overhead due to non-contiguous memory addresses. We believe this MemCpy overhead can be mitigated or eliminated by implementing customized kernels that simultaneously read from and write to non-contiguous memory addresses. Given current impact of this extra MemCpy is minimal, we defer its optimization to future work. 



In practice, Domino achieves up to 50\% to 70\% communication hiding by employing column-wise split on weights. Although this overlapping percentage is lower than batch-wise input split (\S~\ref{sec:implement-input}), this approach remains essential, since that batch-split alone results in tensors with narrow shapes that hinder kernel computation efficiency.   

\subsection{Generic Kernel Optimization}
\label{sec:kernel-opt}

Here we discuss generic kernel-level optimizations with CUDA-MultiStream and PyTorch-native compiling techniques.

\subsubsection{MultiStream}
After splitting the computation into smaller units, the computation required for each kernel is significantly reduced compared to the original TP baseline. To increase GPU utilization while reducing sequential kernel launching overhead, we execute independent operations in parallel using multiple CUDA streams.

To obtain a new CUDA stream, one can retrieve it from the CUDA stream pool. However, this method generates an excessive number of new streams and utilizes them in a round-robin fashion, leading to a high overhead from frequent stream switching. To mitigate this, we first initialize and create a fixed number of global streams before execution. Then, we use an index to obtain a specific stream, thereby reducing the overhead associated with stream switching.

\subsubsection{CudaGraph \& Torch.compile}
torch.compile() functionality from PyTorch accelerates code execution by just-in-time (JIT) compiling PyTorch operations into optimized kernels, enabling improved performance with minimal code modifications~\cite{torch-compile}. Many operations from the torch library are employed to construct our modules. 
By fusing distinct operations, we leverage torch.compile() to enhance our computational efficiency. 


After Domino slicing tensor into multiple chunks, the computation needed for each chunk is significantly less than the original baseline, leading to GPU idleness between adjacent operations (i.e., bubble time). The primary reason for bubbles is the computation time for different operations is less than the PyTorch scheduling latency. To address this issue, we employ CudaGraph~\cite{cuda-graph,nv-cuda-graph} to eliminate the gap between adjacent operations, thereby reducing the overall computation time. 
However, commonly-used on-device random number generator (RNG) feature is incompatible with CudaGraph. As a workaround, we utilize a fixed seed instead of random numbers to mimic the behavior of the RNG.

\section{Evaluation}
This section provides detailed evaluation and benchmark results of Domino. We first discuss the model and hardware settings of our experiments (\S~\ref{sec:model-hardware}).After that, we describe baseline and evaluation metrics (\S~\ref{sec:metrics})
Then we evaluate benchmark results on GPT-3 (\S~\ref{sec:eval-gpt3}) and Llama-2 (\S~\ref{sec:eval-llama2}) models in both single-node and multi-node cases. 

\begin{table}
\centering
\begin{tabular}{|c|c|c|} 
 \hline
 {Model}& {Model Size} & {Global Batch Size}\\ 
\hline
  GPT-3 & 2.7B, 6.7B, 13B, 30B & 4 - 64\\
\hline
Llama-2 & 7B, 13B & 4 - 64\\ 
\hline
\end{tabular}
\caption{Model type, model size, batch size configurations.}
\label{tb:model-config}
\end{table}

\subsection{Model and Hardware}
\label{sec:model-hardware}

Before discussion on detailed evaluation results, we first describe models and hardware settings of our experiments.
\subsubsection{Model} 
We focus on evaluation of GPT~\cite{gpt-2,gpt-3,gpt-4} and Llama~\cite{llama,llama-2,llama-3, llama-3-1} model series. More specifically, we conduct our benchmark tests using GPT-3~\cite{gpt-3} model and Llama-2~\cite{llama-2} model with different model sizes. All model configuration details are illustrated in Table~\ref{tb:model-config}. For model size calculation, we follow the equation from Nvidia Megatron team~\cite{deepak-megatron} as Equation~\ref{eq:model-size}, where $h$ refers to the hidden size and $l$ is the number of layers. $seq\_len$ represents sequence length and $vocab$ is vocabulary size. 

\begin{equation}
\label{eq:model-size}
model\_size = (1 + \frac{13}{12*h} + \frac{vocab+seq\_len}{12*h*l}) * 12*l*h^2
\end{equation}

\subsubsection{Hardware} 
We conduct experiments on Nvidia DGX-H100 boxes~\cite{dgx-h100}, each with 8 H100 GPUs. Within each DGX-H100 node, GPUs are connected with NVLink~\cite{nvlink} and NV-Switch~\cite{nvswitch}. Each DGX-H100 node is equipped with 8 Nvidia InfiniBand ConnectX-7 network cards~\cite{nv-connectx7} for cross node communication, which provides an aggregated network bandwidth of 400 GB/s per node. We have three different hardware settings: 1 node, 2 nodes and 4 nodes, which represents both single node and distributed training environments. All nodes run in the same PyTorch environment with NCCL version 2.18 and CUDA version 12.2. 

\subsection{Metrics}
\label{sec:metrics}
Similar to previous arts on computation-communication overlapping~\cite{t3-overlap, moe-overlap, google-overlap}, we report results analysis mainly with overall training iteration time. In Equation~\ref{eq:tflops}, throughput or TFLOPs can be inferred since it is just inversely proportional to iteration time measurement (i.e., $iter\_time$). 
\begin{equation}
\label{eq:tflops}
TFLOPs \propto \frac{1}{iter\_time}
\end{equation}
We believe iteration time represents more thorough and end-to-end results because CPU side execution (e.g., data pre-processing, learning rate adaptation, etc.) is also taken into account, which may not be included in pure TFLOPs measurements on GPUs. Since we only use TP for model partition and each GPU in TP domain shares the same input, our global batch size is equivalent to our micro batch size.

Our baseline is using a stable release of Megatron-LM with two different settings: synchronous (sync) and asynchronous (async), where sync (i.e., \emph{Megatron-LM(sync)}) means all collective operations are blocking calls and async (i.e., \emph{Megatron-LM(async)}) is to enable backward pass only, coarse-grained computation and communication overlapping feature in Megatron-LM~\cite{megatron-lm}. By default, we compare Domino and \emph{Megatron-LM (async)} with the vanilla \emph{Megatron-LM (sync)} as the baseline. 

One thing worth mentioning is that our Domino scheme is mathematically equivalent to vanilla TP solutions like Nvidia Megatron-LM (as proof in \S~\ref{sec:row-split}, ~\ref{sec:col-split}). With fixed random seed and the same learning rate schedule, we monitored through weights \& bias tool~\cite{wandb}, which shows that Domino's loss curve matches with Megatron-LM baseline. For the sake of brevity, we exclude these convergence results here. 


\begin{figure*}[t]
\centering
\begin{subfigure}{0.33\textwidth}
    \includegraphics[width=1\textwidth]{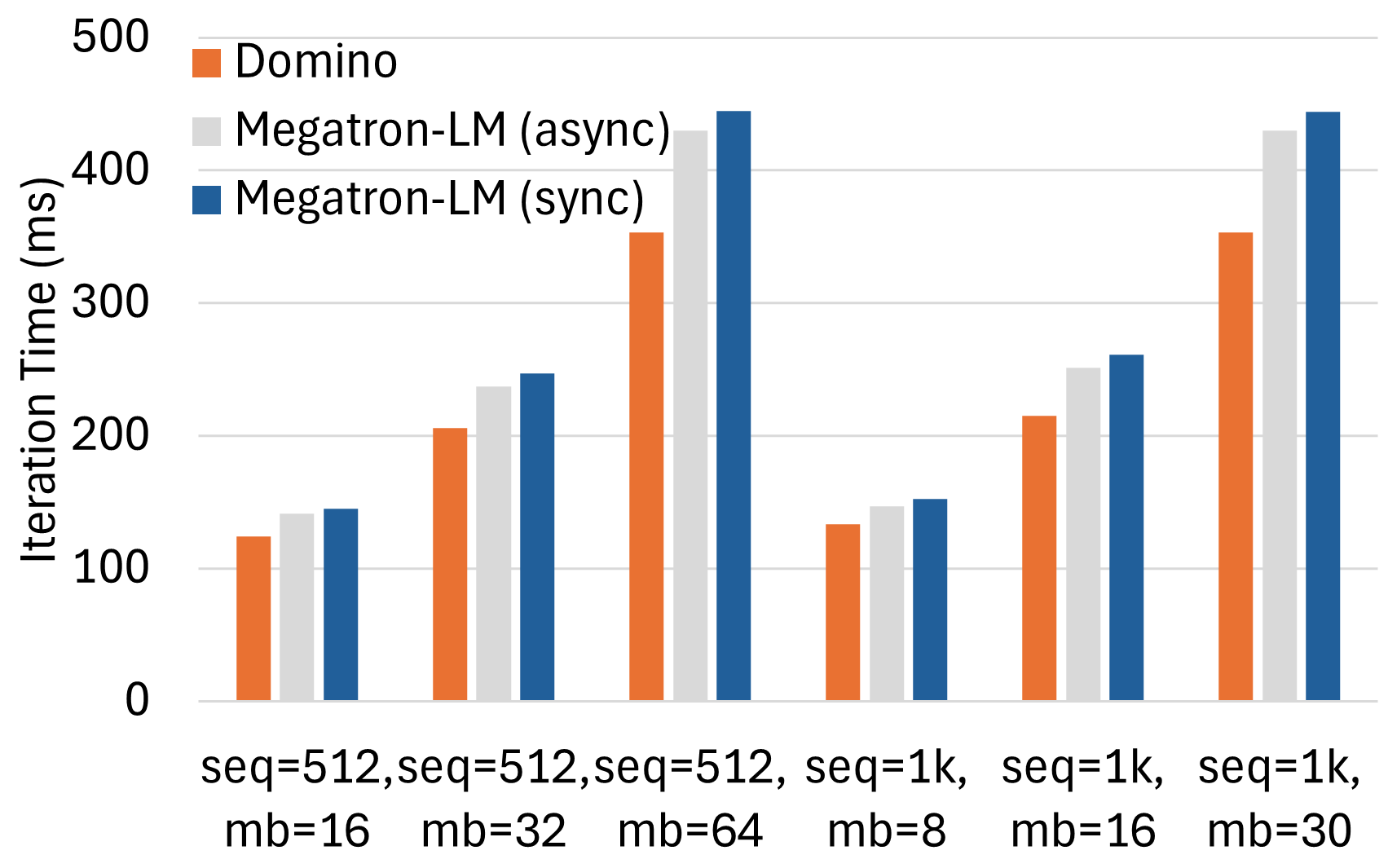}
    \subcaption{GPT-2.7B}\label{fig:gpt-2p7b-1node}
\end{subfigure}
\begin{subfigure}{0.33\textwidth}
    \includegraphics[width=1\textwidth]{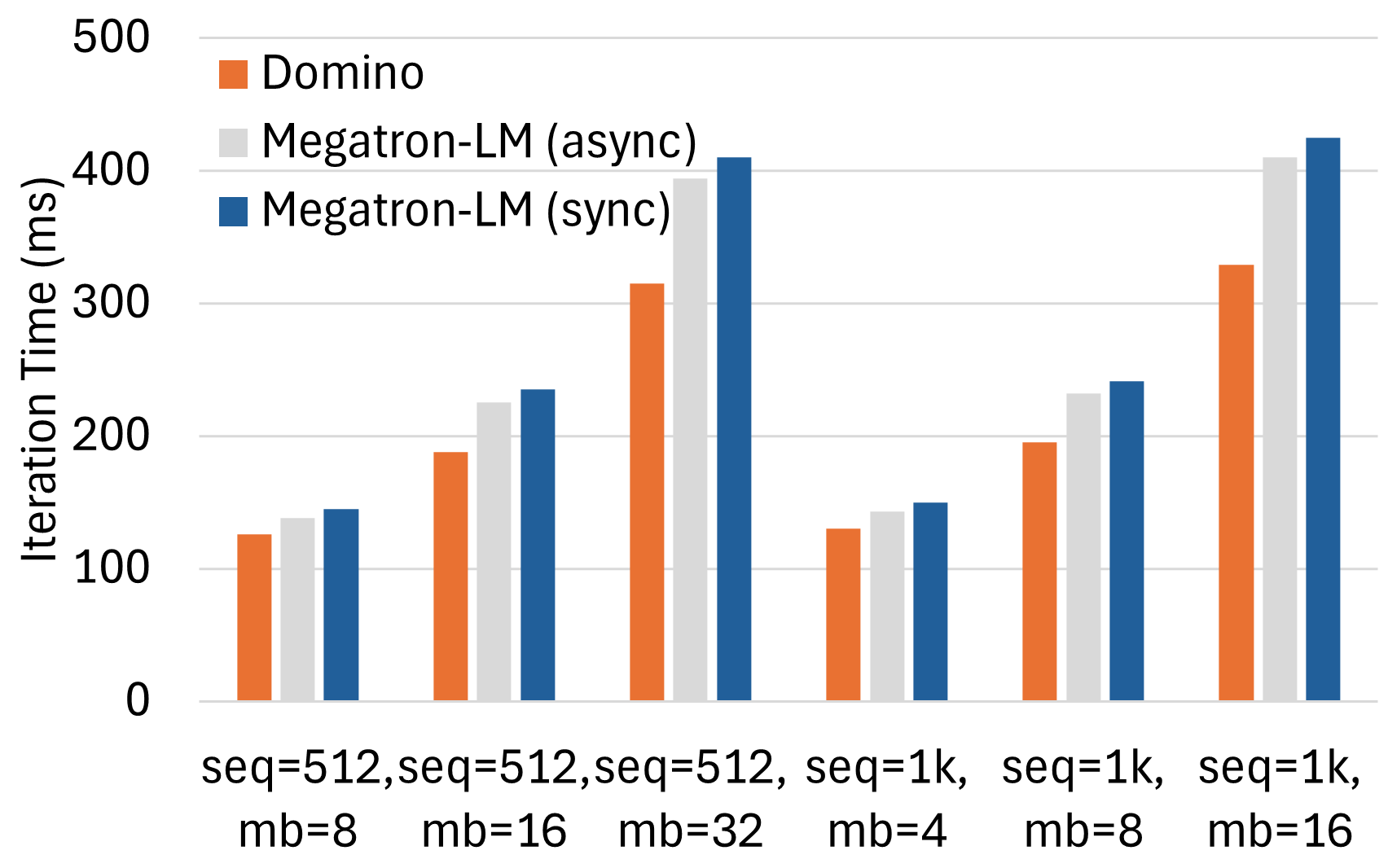}
    \subcaption{GPT-6.7B}\label{fig:gpt-6p7b-1node}
\end{subfigure}
\begin{subfigure}{0.33\textwidth}
    \includegraphics[width=1\textwidth]{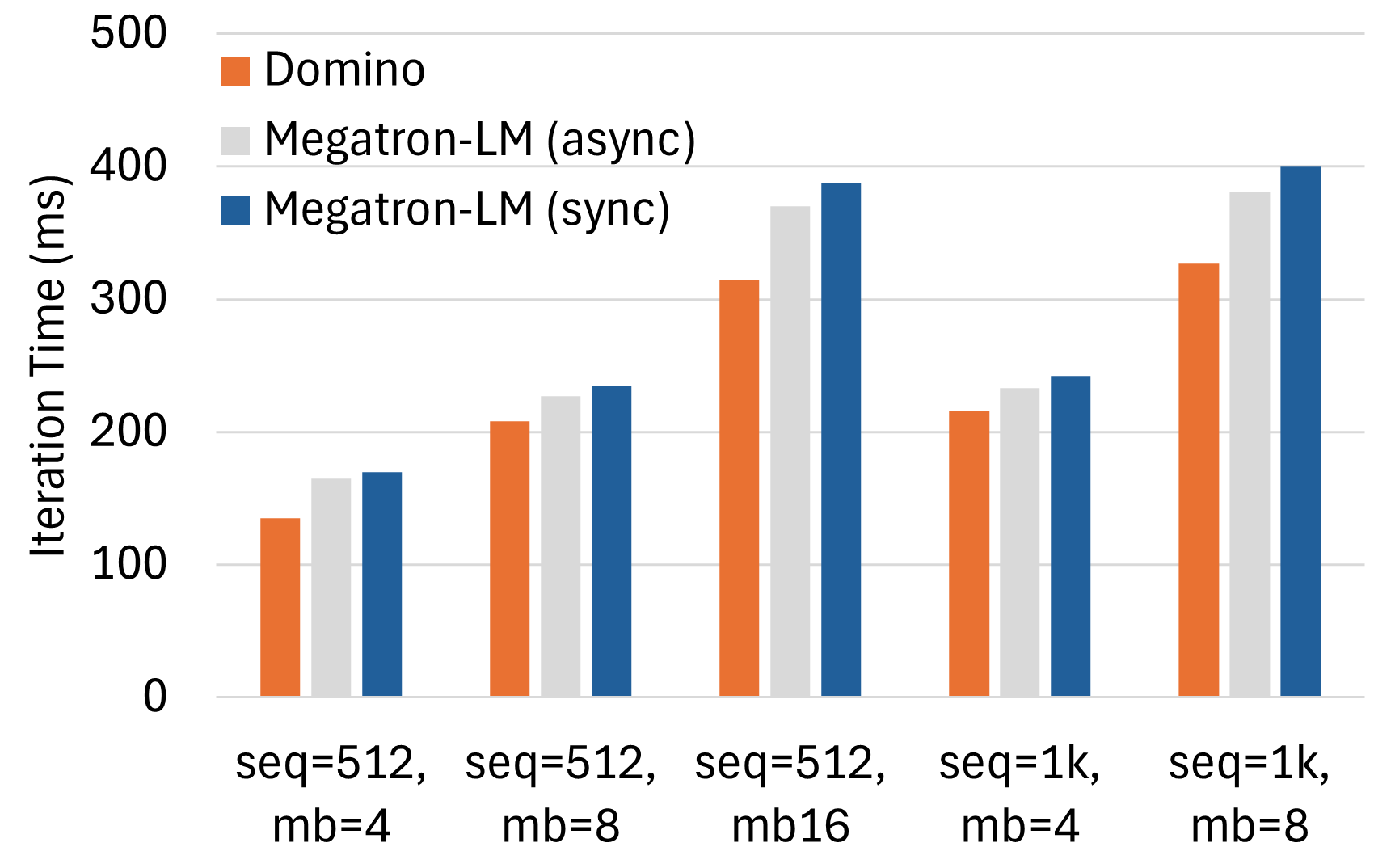}
    \subcaption{GPT-13B}\label{fig:gpt-13b-1node}
\end{subfigure}
\caption{GPT-3 training iteration time with different model sizes, sequence lengths (seq), and micro-batch (mb) sizes on a single DGX-H100 node (8 H100 GPUs).}\label{fig:gpt-1node}
\end{figure*}

\subsection{GPT-3}
\label{sec:eval-gpt3}
GPT is popular and representative model series of transformers. We conduct model training benchmark of GPT-3 with different model sizes ranging from 2.7B to 30B using 1 to 4 DGX-H100 nodes (i.e., 8 to 32 H100 GPUs). We use two different sequence lengths 512 and 1024. Given that for row-split (i.e., batch-split) on input, our smallest micro-batch size starts from 4 and up to the maximum batch size without triggering OOM. We exclude the cases of micro-batch sizes 1 and 2. 
Micro-batch size of 1 is impractical for batch-wise input splitting.
Additionally, micro-batch size of 2 is excluded because with minimum half-half split, each half batch size is 1, which is impossible to achieve good training throughput.

As described in \S~\ref{sec:overview}, we conduct grid search and only report the best performance numbers of Domino via both row-wise input split and column-wise weights split. Additionally, we also tried to enable or disable features like CudaGraph() and torch.compile() and report the best numbers that Domino achieved. 
To achieve good throughput and system efficiency, we report benchmark results of top-2/3 largest micro-batch sizes that $\geq 4$ and without causing OOM.

\subsubsection{Single-node}
\label{sec:gpt-1node}
For single node training, we tested model size with 2.7B, 6.7B and 13B. In summary, compared with Megatron-LM, Domino achieves up to 1.3x throughput speed-up. Furthermore, in multiple cases, Domino achieves near-optimal or even exceeds the optimal settings. The optimal solution refers to disabling all the communication in TP training of both forward and backward passes.

One tricky part is whether we should enable CudaGraph or not. Based on our experimental results, we found that if the batch size is small (i.e., training job is not compute-heavy), enabling CudaGraph could squeeze the bubble/idle time between adjacent kernels thus improving end-to-end performance. On the other hand, if the training job is compute-heavy and does not have much idle time between adjacent kernels, we disable CudaGrpah for faster model training initialization and less on-device memory copy overhead. Taking GPT-3 13B training as an example, with sequence length of 512 and micro-batch size of 4, we notice significant training iteration time reduction (around 10-15\%) if we switch from cudaGraph off mode to on mode, which shows the benefits of reducing idle time between adjacent compute kernels. On the other hand, if we increase the micro-batch size to 16, enabling cudaGraph leads to 5-10\% longer iteration time than disabling it, which is mainly due to extra memory copy overhead introduced in CudaGraph.

\begin{figure}[t]
    \centering
    \includegraphics[width=1\linewidth]{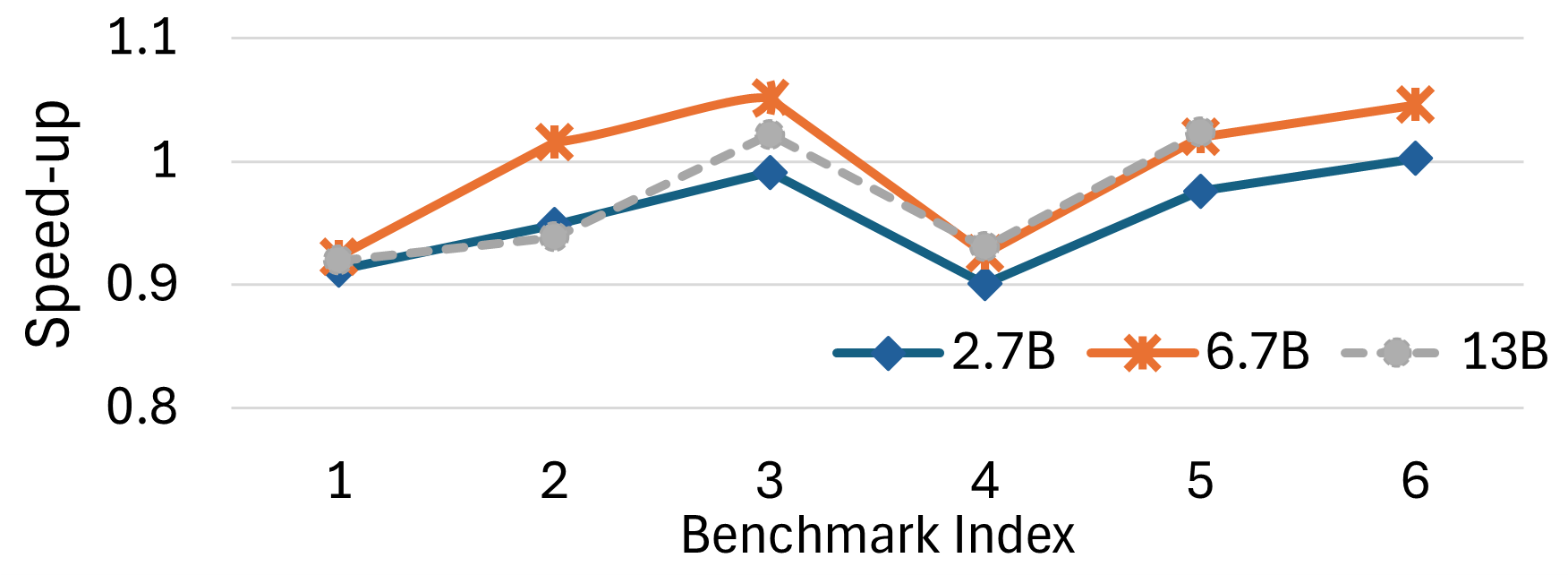}
    \caption{Domino normalized throughput speed-up compared with optimal solution (no communication) on single DGX-H100 (8 H100 GPUs). Benchmark index strictly follows the order of horizontal settings (i.e., seq, mb) in Figure~\ref{fig:gpt-1node}.}
    \label{fig:gpt-1node-optimal}
\end{figure}

Overall, as shown in Figure~\ref{fig:gpt-1node}, enabling Megatron's coarse computation and communication overlapping (i.e., Megatron-LM (async)) achieves around 2-5\% throughput gain compared with vanilla megatron baseline. Compared with the Megatron-LM baseline, Domino achieves higher speedup gains when training batch is large and relatively low performance gains for small batch size cases.  

For GPT-3 2.7B training shown in Figure~\ref{fig:gpt-2p7b-1node}, Domino achieves 1.14x to 1.26x speedup over Megatron baseline for both sequence lengths of 512 and 1k. In GPT-3 6.7B training as Figure~\ref{fig:gpt-6p7b-1node}, since we increase model size from 2.7B to 6.7B, the largest micro-batch sizes are reduced compared with 2.7B cases. However, we achieve the highest throughput gain in 6.7B model compared with 2.7B and 13B cases. More specifically, in Figure~\ref{fig:gpt-6p7b-1node}, for both sequence lengths of 512 and 1k, we achieve from 1.15x to 1.3x speedup over Megatron baseline with increasing micro batch sizes. For 13B cases in Figure~\ref{fig:gpt-13b-1node}, we have the smallest micro-batch sizes for training, which leads to 12\% to 23\% throughput speedup over the Megatron baseline with increased batch sizes. In summary, \emph{Domimo generally outperforms Megatron baseline in varied batch sizes and sequence lengths}. Our performance gain increases as the batch size grows.

\begin{figure*}[t]
\centering
\begin{subfigure}{0.45\textwidth}
    \includegraphics[width=1\textwidth]{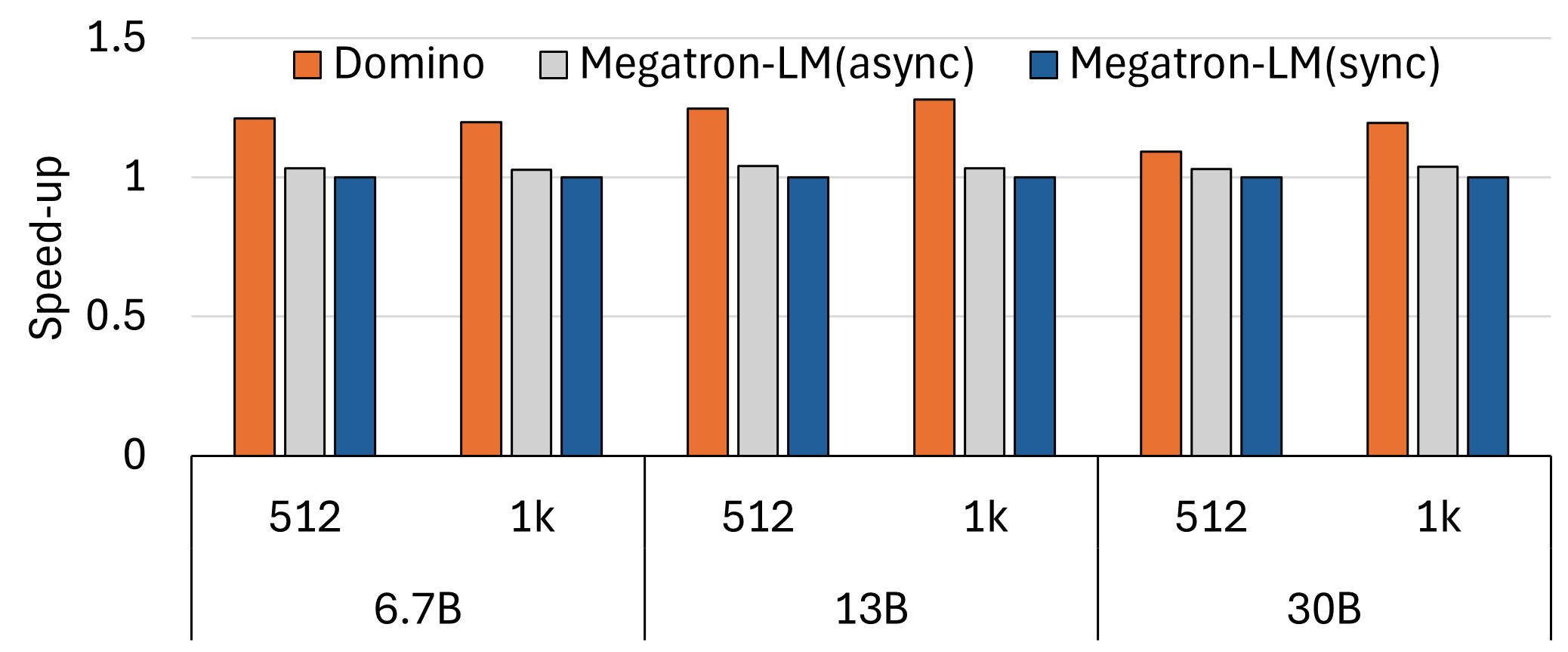}
    \subcaption{2 DGX-H100 (16-H100 GPUs)}\label{fig:gpt-2node-eval}
\end{subfigure}
\begin{subfigure}{0.45\textwidth}
    \includegraphics[width=1\textwidth]{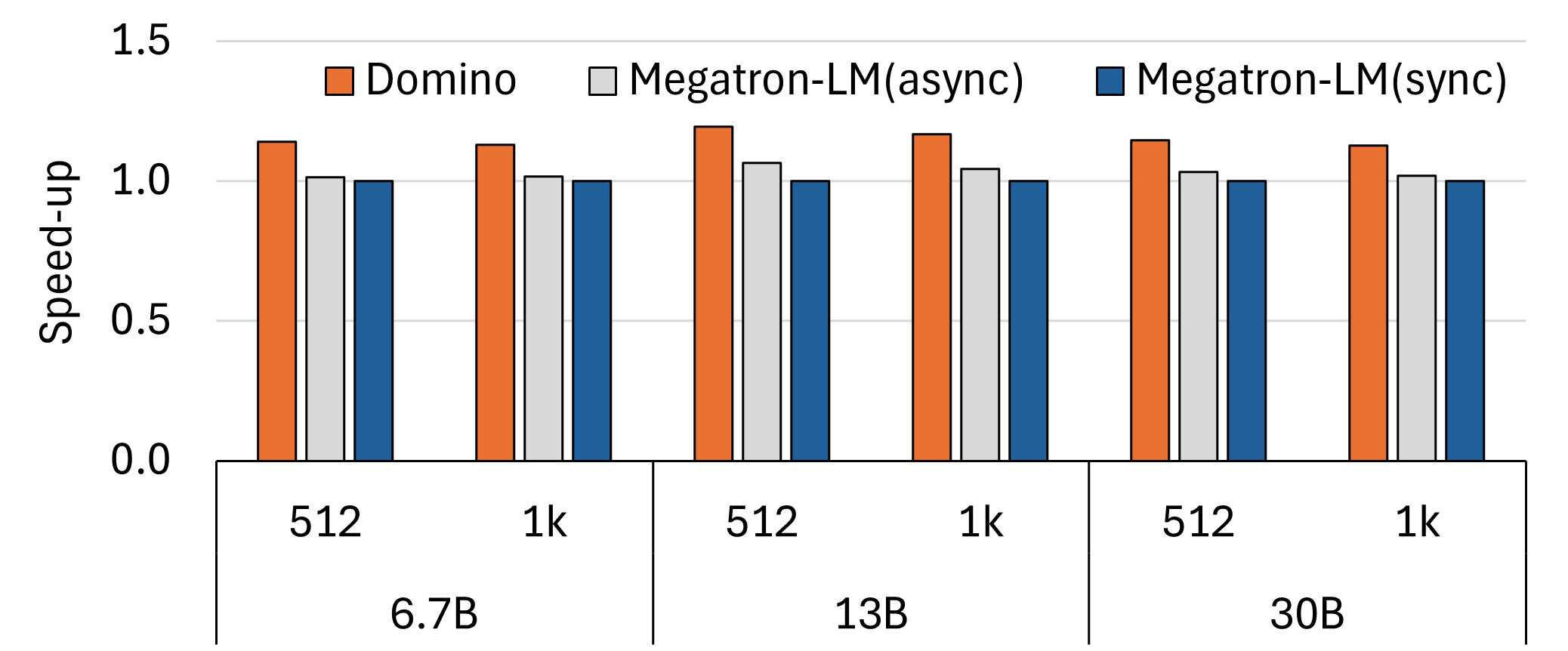}
    \subcaption{4 DGX-H100 (32-H100 GPUs)}\label{fig:gpt-4node-eval}
\end{subfigure}
\caption{GPT-3 training throughput speedup with different model sizes, sequence lengths (seq) and micro-batch (mb) sizes on 2 (16-H100) and 4 DGX-H100 nodes (32-H100 GPUs).}\label{fig:gpt-2-4-node}
\end{figure*}

We also depict comparison of Domino performance with optimal settings with different sequence lengths and batch sizes on a single node. As shown in Figure~\ref{fig:gpt-1node-optimal}, the horizontal benchmark index numbers strictly follow the same order of training settings in Figure~\ref{fig:gpt-1node}. Compared with the optimal setting that removes all communications in Megatron-LM, Domino reaches over 90\% of optimal throughput in all cases and has a few cases even exceeding the optimal settings. We conduct an ablation study and performance gain breakdown.
We find that, for cases where Domino exceeds the optimal setting, the extra performance gain is primarily attributed to our kernel-side optimization as discussed in \S~\ref{sec:kernel-opt}.

\subsubsection{Multi-node}
\label{sec:gpt-4node}

Compared with single node results, multi-node cases are different given cross-node IB bandwidth is still 2-3x lower compared with intra-node NVLink/NVSwitch. Therefore, it is still possible that a single NCCL collective can be longer than the maximum number of computation kernels that Domino can overlap with.

For 2 and 4 DGX-H100 nodes experiments, we test three different model sizes as 6.7B, 13B and 30B across 16 to 32 H100 GPUs with TP model partition strategy. As shown in Fig.~\ref{fig:gpt-2-4-node}, we report normalized throughput speed-up when comparing Domino with Megatron baseline. For both sequence lengths of 512 and 1k, we present our best throughput results with proper batch sizes ranging from 4 to 64. Coarse-grained computation and communication overlapping provided by Megatron-LM (i.e., Megatron-LM (async) in Fig.~\ref{fig:gpt-2-4-node}) gives around 2\%-4\% performance gain on average. 

As shown in Fig.~\ref{fig:gpt-2node-eval}, for 2-node case (16 H100 GPUs), Domino achieves around an average of 1.2x speedup over Megatron baseline for both 6.7B and 30B models with varied sequence lengths and batch sizes. More interestingly, for 13B training, Domino achieves up to 1.3x throughput gain over baseline on 1k sequence length. We believe GPT-3 13B training on 2 DGX-H100 nodes provides a sweet spot that 1) most computation kernels are still highly efficient after our row-wise and column-wise split on inputs and weights, 2) cross-node communication can be mostly overlapped with computation using Domino. 

For 4-node case depicted in Fig.~\ref{fig:gpt-4node-eval}, Domino achieves 1.14x to 1.2x throughput speedup over Megatron baseline across GPT-3-6.7B, GPT-3-13B, GPT-3-30B across different batch sizes and sequence lengths. The reason for less performance gain compared with 2-node cases is cross-node communication cannot be perfectly overlapped with the maximum range of computation kernels of Domino. Given the latest IB with Nvidia ConnectX-8 cards~\cite{nv-connectx8} could provide 800 GB/s inter-node communication bandwidth, we did a simulation projection with 800 GB/s cross-node bandwidth for both Megatron-LM and Domino. In simulation, Domino could achieve up to 1.5x speedup over Megatron baseline. In our simulation, we also note that Domino could potentially achieve higher speedup over Megatron baseline on larger scales (e.g., 128, 256 GPUs). 


\begin{figure}[t]
    \centering
    \includegraphics[width=1\linewidth]{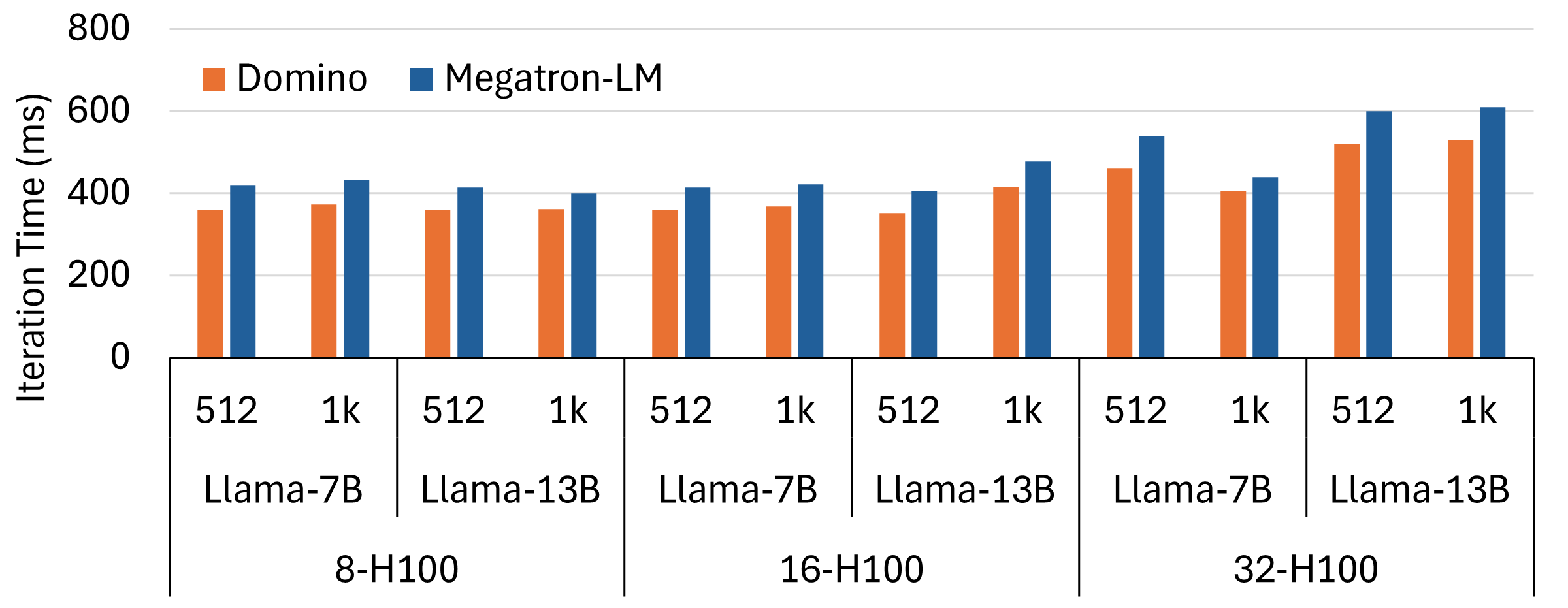}
    \caption{Llama-2 training iteration time with different sequence lengths and model sizes on 1 DGX-H100 node (8-H100), 2 nodes (16-H100), and 4 nodes (32-H100) cases.}
    \label{fig:llama-eval}
\end{figure}

\subsection{Llama-2}
\label{sec:eval-llama2}
We describe our evaluation of Llama-2 model. The major difference between GPT-3 and Llama-2 model is Llama-2 involves new normalization as Root Mean Square Normalization (RMSNorm)\cite{RMSNorm}, new activation function SwiGLU (SwiGated Linear Unit)~\cite{swiGLU} and RoPE (Rotary Position Embedding)~\cite{rope}. Given that the latest Llama-3~\cite{llama-3} model shares a similar model architecture. Compared with Llama-2, some major changes of Llama-3 are RoPE~\cite{rope} size, hidden size, and embedding configuration difference. Therefore, our results on Llama-2 can be representative of the Llama model series. To avoid duplicated results patterns such as \S~\ref{sec:eval-gpt3}, we only report the results of the largest batch size without triggering OOM. Since Megatron with its coarse computation communication overlapping show similar performance as vanilla baseline (only 2\% to 4\% gain on average), we exclude the Megatron-LM (async) result here. Similar to \S~\ref{sec:eval-gpt3}, we also benchmark two different sequence lengths as 512 and 1024 (i.e., 1k in Fig.~\ref{fig:llama-eval}). 

\subsubsection{Single-node}
For single node experiments as shown in left-most 8 bars under 8-H100 column in Figure~\ref{fig:llama-eval}, Domino achieves around 1.16x speedup for Llama-7B training, and 1.1 to 1.15x speedup for Llama-13B training. The lower performance gain on larger model is because we can support smaller batch size training. With smaller batches, Domino's kernel launching overhead becomes more noticeable thus leading to less throughput gain. 

Compared with results from GPT-3, Domino has less performance gain over Megatron-LM. The main issue is due to the rotary embedding feature introduced in Llama-2 model. This rotary embedding builds extra data dependency among our input batch dimension split pieces. For better system performance, we leave this rotary embedding issue as a future optimization direction. 


\begin{figure}[t]
    \centering
    \includegraphics[width=1\linewidth]{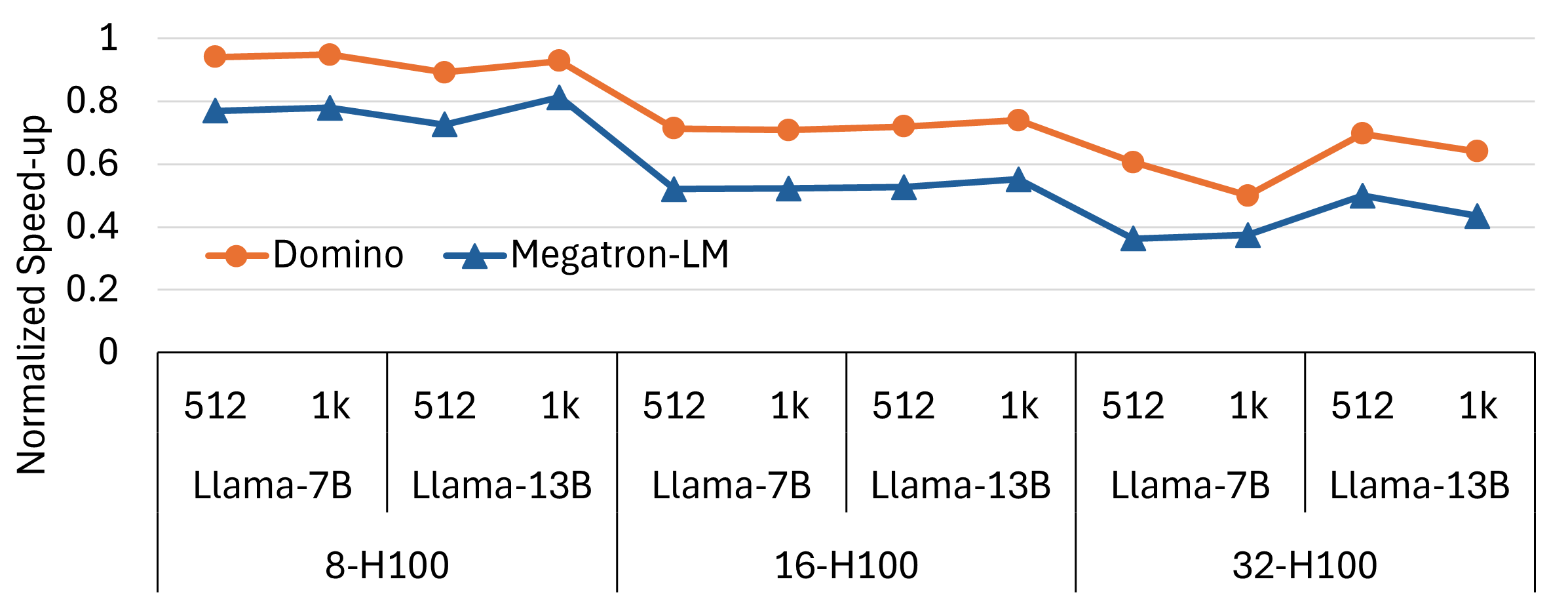}
    \caption{Normalized speedup when comparing Domino, Megatron with Optimal setting (i.e., no comm.) for Llama-2 training on 1 DGX-H100 node (8-H100), 2 nodes (16-H100) and 4 nodes (32-H100) cases.}
    \label{fig:llama-optimal}
\end{figure}

We also compare both Domino and Megatron-LM with optimal throughput scenarios (i.e., no communication). As shown in Figure~\ref{fig:llama-optimal}, the left most 4 groups of data points under 8-H100 column present Domino and Megatron-LM throughput performance. Here we normalize optimal throughput as 1 and calculate the corresponding throughput of both Domino and Megatron. For Domino in 8-H100 cases, we mostly achieve around 90+\% optimal throughput and 10\% better than Megatron, which is quite decent.

\subsubsection{Multi-node}
Similar to single node training, we benchmark Llama-2 7B and 13B training on 2 and 4 nodes with sequence lengths of 512 and 1k. 

In Figure~\ref{fig:llama-eval}, two node cases are the middle 8 bars under 16-H100 column. Compared with Megatron-LM, Domino achieves around 1.15x speedup for both 512 and 1k sequence lengths. For 4 nodes case, similar results are shown as the right-most 8 bars under 32-H100 column of Figure~\ref{fig:llama-eval}. Domino achieves 1.08x to 1.17x speedup over Megatron-LM in various model sizes and sequence lengths of 4-node cases.

When comparing to the optimal case (i.e., no communication), as shown in Fig.~\ref{fig:llama-optimal}, for both 2-node and 4-node cases, Domino achieves around 60-80\% of optimal throughput and consistently around 10-20\% better than Megatron-LM.

\section{Related Work}
\label{sec:related}
Previous literature on reducing communication overhead in distributed model training mainly falls into two categories: overlapping communication with computation, and optimization on collective communication.
\subsection{Overlapping Communication with Computation}
\label{sec:related-overlap}
One major line of overlapping communication with computation is to provide a better scheduling policy. Centauri~\cite{centauri-overlap} is a recent work on overlapping communication and computation for hybrid parallelism (e.g. DP, TP, PP, SP) scenarios. Its multi-level partitioning and scheduling introduce significant planning overhead. Additionally, the generated schedule is complex which makes it hard for end-to-end correctness debugging, thus making the proposed scheme less practical. Furthermore, adopting hierarchical collective (e.g., all-gather) does not reduce the overall cross-node communication volumes, and hierarchical collective calls (first intra-node then inter-node) could lead to longer end-to-end network latency in practice. 
Alpa~\cite{alpa_osdi22} conducts compiler-level optimization on intra and inter-operator parallelism as well as better overlapping with communication. Similar as Centauri, Alpa's tensor partition strategy is complicated and at compiler-level, which makes it almost impossible to conduct correctness debugging on the user side. Different from Alpa, Domino is mainly at kernel scheduler-level and our solution is clean and neat for correctness debugging. Furthermore, Alpa achieves similar throughput as Megatron-LM while Domino outperforms Megatron-LM.  
Lancet~\cite{moe-overlap} leverages unique feature of MoE (Mixture-of-Experts) model and overlap All-to-all collectives with forward and backward computation, which is orthogonal to Domino as we focus on dense model type as it is more widely used (e.g. Llama~\cite{llama,llama-2,llama-3, llama-3-1}, GPT~\cite{gpt-3, gpt-4}, Phi~\cite{phi3} model series).
TicTac~\cite{tictac-overlap} provides near-optimal computation communication overlapping in Parameter Server (PS)~\cite{ps_osdi14} architecture. However, TicTac approach cannot be applied in LLM training since modern large model training only uses MPI-based architecture (i.e. all-worker)~\cite{mpi,pathways} rather than PS. 
Breadth-first pipeline-parallism~\cite{breadth-pp} partitions model layers to GPUs in a round-robin~\cite{wavelet} fashion, which interleaves computation with communication and mitigates the burst of communication calls in vanilla pipeline parallel training. This optimized pipeline parallelism is beneficial for low bandwidth interconnect scenarios. However, it may have minor performance gain for state-of-the-art HPC (High-Performance Computing) clusters equipped with high bandwidth InfiniBand~\cite{infiniband} links as the scenarios that Domino focuses on. 

Another main line of work is kernel fusion of computation and communication. Researchers from Google~\cite{google-overlap} focus on intra-layer overlapping via kernel fusion of GeMM with collective operations on TPUs. T3~\cite{t3-overlap} and Flux~\cite{flux-bytedance} apply and extend similar ideas on Nvidia GPUs. However, these kernel fusion works only overlap collectively with one specific compute kernel type (i.e., GeMM), which limits its overlapping scope. 
CoCoNet~\cite{coconet-overlap} provides a general kernel fusion paradigm that automatically generates fused kernel between collective and popular compute operations (i.e., GeMM and convolution). However, the generated code achieves less system efficiency compared with directly using highly-optimized kernels from cuBlas~\cite{cublas}, cutlass~\cite{cutlass} or cuDNN~\cite{cudnn} mainly due to extra in-kernel synchronization introduced for this fine-grained compute-communication overlapping. MGG~\cite{mgg} fuses computation and communication kernels for graph neural network (GNN) via NVSHMEM~\cite{nvshmem}. Researchers from AMD~\cite{superOP-amd} also fuse embedding and GeMM with collectives to achieve fine-grained compute-communication overlapping on AMD MI200-series GPUs.
Compared with these GeMM+NCCL fusion work, Domino provides more flexible and wider range of computation and communication overlapping.  
Furthermore, Domino is orthogonal to this compute-communication kernel fusion line of work, which can be adopted to further improve Domino system efficiency. 

\subsection{Fast Collective Communication}
\label{sec:related-nccl}
Collective communication libraries like NCCL~\cite{nccl}, Gloo~\cite{gloo}, Blink~\cite{blink}, Horovod~\cite{horovod}, optimize collective communication itself to reduce communication overheads in distributed model training. For example, NCCL~\cite{nccl} incorporates Infini-band Sharp technology~\cite{ib-sharp} and its CollNet optimization~\cite{nccl-collnet} for in-network data aggregation to reduce communication volume as well as tensor reduction overhead. ACE~\cite{ace} is a similar work of offloading collectives into network fabric from academia, which provides good simulation numbers.  Gloo~\cite{gloo} provides general collective primitive supports on CPU side. Horovod~\cite{horovod} reduces communication overhead by batching multiple collective calls, thus reducing kernel launching overheads. Blink~\cite{blink} improves network utilization by providing a spanning tree communication protocol that leverages idle links that cannot form ring topology. ByteScheduler~\cite{byte_scheduler} incorporates both parameter server architecture~\cite{ps_osdi14} and MPI all-worker architecture~\cite{mpi,mpi-opt,mpitutorial} for hybrid collective primitive design, and switches between these two schemes for better network utilization. MSCCL~\cite{msccl} and its following work~\cite{mscclpp, taccl} optimize collective communication via various technologies such as compiler optimization~\cite{msccl,mscclpp}, sketch abstraction~\cite{taccl}, etc.

Domino is orthogonal to all collective optimizations and can plug in any collective library if needed. We choose NCCL by default given its wide adoption in distributed model training on Nvidia GPUs~\cite{dgx-h100,dgx-a100}. To enable Domino to run on AMD GPUs, simply replacing NCCL calls with corresponding RCCL~\cite{rccl} collectives from AMD would work seamlessly. 
\section{Conclusion}

We propose Domino, a generic approach for tensor slicing and partitioning to achieve fine-grained overlapping of computation kernel sequences with communication collectives. Extensive results on multiple DGX-H100 nodes show that, Domino can achieve up to 1.3x speedup over the state-of-the-art tensor parallelism solution as Megatron-LM. Furthermore, Domino even exceeds optimal performance (i.e., remove all communication in Megatron-LM) in some cases. With the trend of high communication bandwidth and faster computation per accelerator, Domino could be beneficial for both small scale and large-scale LLMs training.


\section*{Acknowledgments}

We would like to thank Shiyang Chen (Rutgers University), Shuaiwen Leon Song (together.ai), Hexu Zhao (NYU) and Yuxiong He (Snowflake) for valuable feedbacks and code contributions to our Domino project. 

\section*{Availability}

Domino is an open source project from Microsoft DeepSpeed Team (\url{https://www.deepspeed.ai/}). Code implementation will be publicly available at \url{https://github.com/microsoft/DeepSpeed}

\clearpage
\bibliography{domino}
\bibliographystyle{plain}

\end{document}